\documentstyle[aps,psfig,preprint]{revtex}
\tighten
\begin{document}
\newcommand{\be}{\begin{equation}}
\newcommand{\ee}{\end{equation}}
\newcommand{\bea}{\begin{eqnarray}}
\newcommand{\eea}{\end{eqnarray}}
\newcommand{\f}{\frac}  
\newcommand{\la}{\lambda}
\newcommand{\ve}{\varepsilon}
\newcommand{\ep}{\epsilon}
\newcommand{\da}{\downarrow}
\newcommand{\V}{{\cal V}}
\newcommand{\ovl}{\overline}
\newcommand{\Ga}{\Gamma}
\newcommand{\ga}{\gamma}
\newcommand{\bra}{\langle}
\newcommand{\ket}{\rangle}
\title{Attenuation of the intensity within a superdeformed band}
\date{August 29, 2001}
\author{A.J. Sargeant$^1$, M. S. Hussein$^1$, M. P. Pato$^1$, N. Takigawa$^2$ 
and M. Ueda$^1$}
\address{$^1$Nuclear Theory and Elementary Particle Phenomenology 
Group, Instituto de F\'\i sica, Universidade de S\~{a}o Paulo,
Caixa Postal 66318, 05315-970 S\~{a}o Paulo, SP, Brazil}
\address{$^2$Department of Physics, Tohoku University, Sendai, 980-8578, Japan}
\maketitle
\begin{abstract}
The attenuation of the intra-band intensity of a superdeformed band
which results from mixing with normally deformed configurations is 
calculated using reaction theory. It is found that the sharp increase of the
attenuation is mostly due to the tunnelling through a spin dependent barrier
and not to the chaotic nature of the normally deformed states.
\end{abstract}
It is now well established that the intensities of $E$2 gamma transitions 
within a superdeformed (SD) rotational band show cascades down to low angular 
momentum \cite{Tw 86,At 93,Kh 93,Kr 94,Kr 96,Ku 97,Kr 97}. These cascades 
exhibit the distinct feature that the intensity remains constant until a 
certain spin is reached where-after the intensity drops to
zero within a few transitions. 
The sharp drop in intensity is
commonly referred to as the decay out of a superdeformed band and is believed 
to arise from mixing of the SD states with normally deformed (ND) states 
of identical spin. 

The earliest theoretical work to implement such an interpretation 
\cite{Vi 90b,Vi 90a,Sh 92,Sh 93} used a statistical model of the coupling
between the SD and ND states.  More recently, Refs. 
\cite{We 98,Gu 99} used a framework originally developed for
the study of compound nuclear reactions to derive formulae for the intensity 
in a more rigorous fashion (the expressions for the intensity in Refs. 
\cite{Vi 90b,Vi 90a,Sh 92,Sh 93} are deduced from probability arguments). 
Ref. \cite{Gu 99} concluded that Refs. \cite{Vi 90b,Vi 90a,Sh 92,Sh 93} 
are valid in the non-overlapping resonance region.
Refs. \cite{Vi 90b,Vi 90a,Sh 92,Sh 93} further calculate the spin 
dependence of the relevant parameters
(the electromagnetic widths of the SD and ND states, the level density
of the ND states and spin dependence of the barrier separating the
SD and ND wells)
which Refs. \cite{We 98,Gu 99} do not.
Two features common to Refs. \cite{Vi 90b,Vi 90a,Sh 92,Sh 93,We 98,Gu 99} 
are (i) the use of
the Gaussian Orthogonal Ensemble (GOE) to simulate the ND states 
(ii) the use of the ``golden rule'' to extract a
width for the the SD states due to mixing with the ND states.

Here, as in Refs. \cite{We 98,Gu 99} we exploit the similarity between  
the decay out of superdeformed bands and compound nuclear reactions to 
write the intensity as the sum of average and fluctuation contributions. 
However we use an energy average in place of the ensemble average used in
Refs. \cite{We 98,Gu 99}. The energy average approach allows the inclusion
of the following features which are more difficult to incorporate into an 
ensemble average. 
\\(i) A hierarchy of complexity in the ND spectrum may be introduced.
\\(ii) A statistical model different from the GOE may be used to simulate
the ND states, as was proposed in Refs. \cite{Ab 99,Ab 99b}.
\\(iii) A width for the SD states due to mixing with the ND states arises
naturally without appealing to the ``golden rule'' whose range of validity
has been found to be restricted \cite{Fr 96,Sa 99}. 
   
\begin{figure}[ht]
\centerline{\psfig{figure=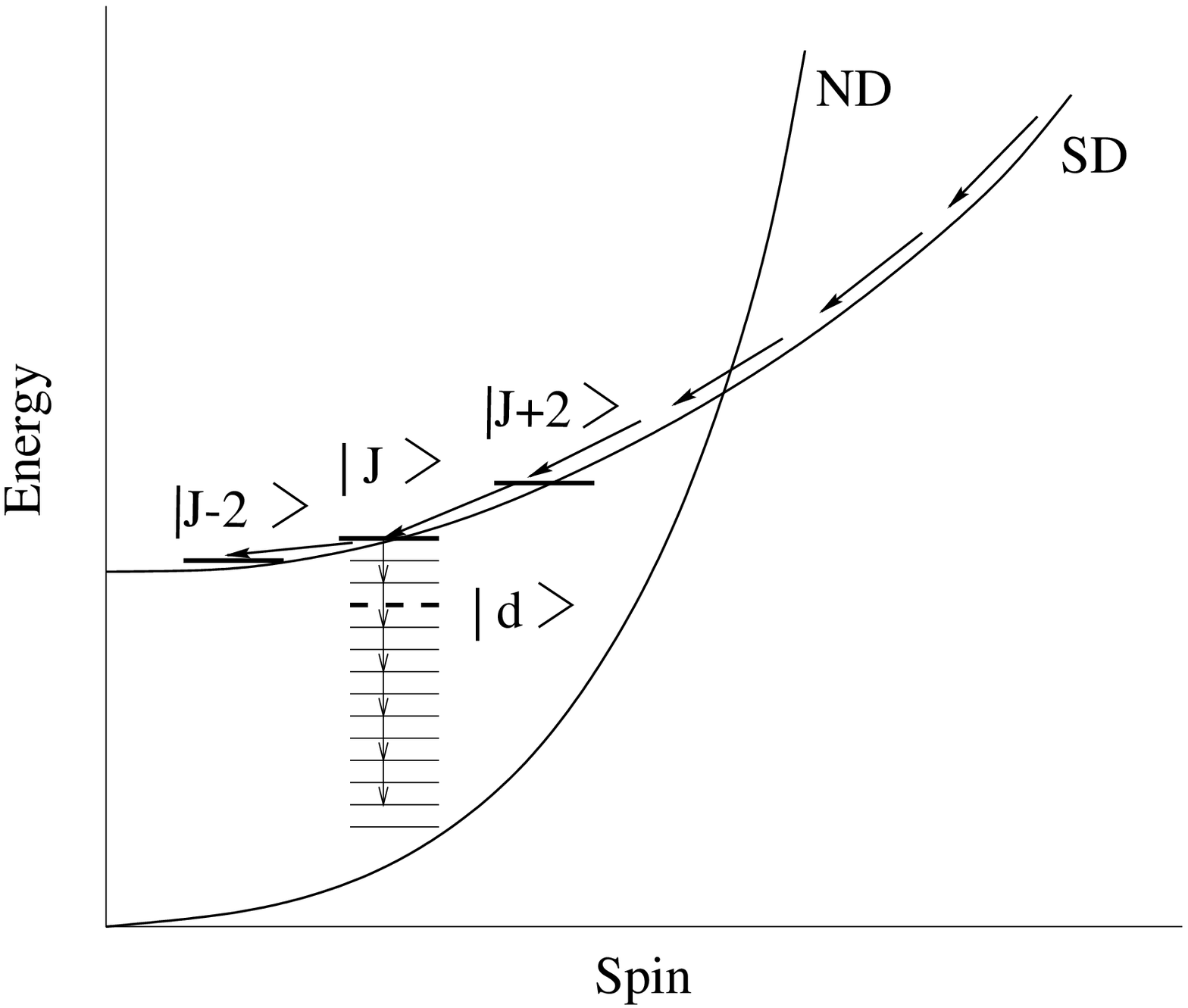}}
\caption{Schematic diagram illustrating the decay out of a superdeformed band.}
\label{schdi}
\end{figure}

In Figure \ref{schdi} we show a schematic plot of the energy of ND and 
SD bands as a function of spin. The observable in which
attenuation is seen is the total intensity of two consecutive $E$2 photons 
in the cascade down the SD band. Let $|J\rangle$ denote an SD configuration 
with spin $J$. The relative intensity of the two step transition
$|J+2\rangle\stackrel{\gamma_1}{\rightarrow}
|J\rangle\stackrel{\gamma_2}\rightarrow|J-2\rangle$ 
(relative to the intensity of the same two step transition
in the absence of mixing with other configurations) is given by 
\bea
\nonumber
F_J&=&\f{1}{2\pi\Gamma_{J+2}^{\gamma}}\int_{-\infty}^{\infty}dE_{\gamma_1}
\int_{-\infty}^{\infty}dE_{\gamma_2}|
\bra J-2|T(E_{J-2}+E_{\gamma_2})|J+2\ket|^2 
\delta(E-E_{J-2}-E_{\gamma_2})
\\&=&\f{1}{2\pi\Gamma_{J+2}^{\gamma}}\int_{-\infty}^{\infty}
dE|\bra J-2|T(E)|J+2\ket|^2,
\label{0F}
\eea
where $E\equiv E_{J+2}-E_{\gamma_1}=E_J$ takes account of the Hamiltonian 
of the electromagnetic field, $E_J$ being the 
energy of $|J\rangle$ and $E_{\gamma_1}$ and $E_{\gamma_2}$ the energies
of the two consecutive photons. The electromagnetic width
of $|J+2\rangle$ is $\Gamma_{J+2}^{\gamma}$
making $2\pi\Gamma_{J+2}^{\gamma}$ the intensity when there is no
mixing with the ND states and thus no flux loss from the SD band. 
Note that in Eq. (\ref{0F}) we ignore the widths of the initial 
and final states for the purpose of calculating the relative intensity.

The transition amplitude is given by
\be
\bra J-2|T(E)|J+2\ket=\gamma_{J+2}\langle J|G(E)|J\rangle\gamma_{J}.
\label{T}
\ee
Here $\gamma_{J+2}$ is the electromagnetic decay amplitude of $|J+2\rangle$ 
defined such that $\Gamma_{J+2}^{\gamma}=\gamma_{J+2}^2$; $\gamma_{J}$ and
$\Gamma_{J}^{\gamma}=\gamma_{J}^2$ are the corresponding decay amplitude and 
width of $|J\rangle$ whilst
the Green's function is given by
\be
G(E)=(E-H)^{-1}.
\label{0G}
\ee
The total nuclear Hamiltonian, which takes the coupling 
to the electromagnetic field into account, is denoted by H. 

The projected Green's function $\langle J|G(E)|J\rangle$ 
may be expressed in terms of
its Lorentzian energy average 
$\langle J|G^{\rm{av.}}(E)|J\rangle=\langle J|G(E+\f{iI}{2})|J\rangle$ 
(energy averaging interval $I$)
plus a fluctuation part \cite{Ka 73,Fe 92}:
\be
\langle J|G|J\rangle=\langle J|G^{\rm{av.}}|J\rangle+
\langle J|G^{\rm{fluc.}}|J\rangle,
\label{avfluc}
\ee
where by definition the energy average of $\langle J|G^{\rm{fluc.}}|J\rangle$ 
is zero.
Thus Eq. (\ref{0F}) for the relative intensity may be written
\be
F_J=F_J^{\rm{av.}}+F_J^{\rm{fluc.}},
\label{Favfluc}
\ee
where
\be
F_J^{\rm{av.}}=\f{\Gamma^{\gamma}_{J}}{2\pi}
\int_{-\infty}^{\infty}dE
|\langle J|G^{\rm{av.}}|J\rangle|^2
\label{Fav}
\ee
and
\be
F_J^{\rm{fluc.}}=\f{\Gamma^{\gamma}_{J}}{2\pi}
\int_{-\infty}^{\infty}dE
\left(|\langle J|G^{\rm{fluc.}}|J\rangle|^2\right)^{\rm{av.}}.
\label{Ffl}
\ee
In this paper we focus our discussion on $F_J^{\rm{av.}}$. It can be shown 
that
\be
\langle J|G^{\rm{av.}}(E)|J\rangle=
\f{1}{E-E_J+i\Gamma^{\gamma}_{J}/2-W_{JJ}(E)}.
\label{GavJJ}
\ee
The derivation of Eq. (\ref{GavJJ}) for 
$\langle J|G^{\rm{av.}}|J\rangle$ and an expression for
$\langle J|G^{\rm{fluc.}}|J\rangle$ using projection operator techniques
will be reported in a subsequent paper.

The form of $W_{JJ}(E)$ depends on the specific model for the Hamiltonian
which is employed. It is our aim 
to study whether or not the chaotic nature (as classified by 
random matrix theory [RMT]) of the ND states
is decisive in explaining the observed 
attenuation. In order to isolate the statistical aspects of the calculation 
we use two different
models distinguished by whether the tunnelling interaction 
mixes $|J\rangle$ randomly with the ND states (\emph{model A})
or whether it couples more strongly to certain ND states 
than others (\emph {model B}). In the latter we shall make the most extreme 
assumption that $|J\rangle$ couples to only one ND state. 

\emph{Model A} is represented by the matrix
\be
H\to\left(
\begin{array}{cc}
E_{J}& V_{Jn}
\\V_{Jn}         &  E_{n}\delta_{n'n}
\end{array}
\right)
-\f{i}{2}
\left(
\begin{array}{cc}
\Ga^{\ga}_J   & 0
\\0                &   \Ga^{\ga}_N\delta_{n'n}
\end{array}
\right),
\hspace{.5cm}n=1,...,N,
\label{Weidmatrix}
\ee
where $E_n$ denotes the energies of the $N$ ND states with which $|J\rangle$
mixes due to the real tunnelling interaction $V_{Jn}$. Here $\Ga^{\ga}_N$
is an electromagnetic width which we assume to be common to the ND states.
With these definitions $W_{JJ}(E)$ becomes
\be
\label{WJJA}
W_{JJ}(E)=\sum_{n=1}^N\f{\left[V_{Jn}\right]^2}
{E-E_n+i\left(\Ga^{\ga}_N+I\right)/2}.
\ee

The energies $E_n$ are constructed using the
deformed Gaussian orthogonal ensemble (DGOE) \cite{Hu 93}. 
The DGOE allows a smooth interpolation from Poisson to GOE statistics
by varying a mixing parameter $\la$ from 0 to 1. Thus the $E_n$ are 
the eigenvalues of a random Hamiltonian $h$ which is real
symmetric and whose matrix elements are taken to be Gaussian distributed random
numbers with zero mean and variances
\be
\bra h^2_{nn}\ket=\f{2}{N},\hspace{.3cm}
\bra h^2_{n'n}\ket=\f{\la^2}{N},n'\ne n.
\ee
The random tunnelling interaction is taken to have zero mean and variance
\be
\bra V^2_{Jn}\ket=v^2_J.
\ee
We assume that $E_J$ lies in the middle of the $N$ ND states, that is,
$E_J=0$. Following Refs. \cite{Vi 90b,Vi 90a,We 98,Gu 99} we introduce
a spreading width $\Ga^{\da}_J$ through the 
golden rule 
\be\label{golden}
\Ga^{\da}_J=2\pi v^2_J.
\ee

We maintain doubts about the meaningfulness of 
Eq. (\ref{golden}) regarding its interpretation as a width \cite{Sa 99}. 
For practical purposes, however,
it is a change of variable from $v_J$ to $\Gamma_J^{\downarrow}$.
All quantities of dimensions energy
are to be understood to have units of $D_J$, where $D_J$ denotes
the mean spacing of the ND states around $E_J$. 
Thus the $E_n$ (and the $E_q$ and $\V_{dq}$ in 
Eq. (\ref{ourmatrix})) which are generated from the DGOE are to be understood
to have been unfolded such that the $E_n$ and $E_q$ have mean 
spacing equal to unity.
Thus we may write $W_{JJ}(E)$ as
\be
\label{2WJJA}
W_{JJ}(E)=\f{\Ga^{\da}_J}{2\pi}\sum_{n=1}^N\f{g_n^2}
{E-E_n+i\left(\Ga^{\ga}_N+I\right)/2}.
\ee
where the $g_n$, $n=1,..,N$, are Gaussian distributed random numbers with
zero mean and unit variance.

\emph{Model B} is represented by the matrix
\be
H\to\left(
\begin{array}{ccc}
E_{J}    &  V_{Jd}           &   0          \\
V_{Jd} &  E_{d}           &  {\cal V}_{dq} \\
0        & {\cal V}_{dq}  &  E_{q}\delta_{q'q}
\end{array}
\right)
-\f{i}{2}
\left(
\begin{array}{ccc}
\Ga^{\ga}_J   & 0                    & 0          \\
0                &  \Ga^{\ga}_N      & 0            \\

0                &  0                    & \Ga^{\ga}_N\delta_{q'q}
\end{array}
\right),
\hspace{.5cm}q=1,...,N-1.
\label{ourmatrix}
\ee
We assume here that $|J\ket$ couples to only one state of normal 
deformation; $|d\ket$, which has energy $E_d$; with strength $V_{Jd}$. 
This special
state is subsequently mixed with other ND configurations with energies 
$E_q$ by a residual interaction $\V_{dq}$.
Now $W_{JJ}(E)$ becomes
\be
\label{WJJB}
W_{JJ}(E)=\left[V_{Jd}\right]^2\sum_{n=1}^N\f{\left[c_d(n)\right]^2}
{E-E_n+i\left(\Ga^{\ga}_N+I\right)/2}
\ee
where $c_d(n)$ denotes component $d$ of the $n$th eigenvector 
of the sub-matrix of the first term of Eq. (\ref{ourmatrix})
obtained by excluding the first row and the first
column. Now the $E_q$ are eigenvalues
of a random Hamiltonian $h$ which is real
symmetric and whose matrix elements are taken to be Gaussian distributed random
numbers with zero mean and variances 
\be
\bra h^2_{qq}\ket=\f{2}{N},\hspace{.3cm}
\bra h^2_{q'q}\ket=\f{\la^2}{N},q'\ne q.
\ee
The residual interaction is also taken to have zero mean and variance
\be
\bra \V^2_{dq}\ket=\f{\la^2}{N}.
\ee
We put $E_d=E_J=0$. Introducing 
\be\label{golden2}
\Ga^{\da}_J=2\pi\f{\left[V_{Jd}\right]^2}{N},
\ee
we can write Eq. (\ref{WJJB}) as
\be
\label{2WJJB}
W_{JJ}(E)=\f{\Ga^{\da}_J}{2\pi}\sum_{n=1}^N\f{N\left[c_d(n)\right]^2}
{E-E_n+i\left(\Ga^{\ga}_N+I\right)/2}.
\ee
Thus comparing Eq. (\ref{2WJJA}) with Eq. (\ref{2WJJB}) we see that,
although the meaning of $\Ga^{\da}_J$ is different for the two models,
the difference between \emph{model A} and \emph{model B} boils down
how much the distribution $\left[c_d(n)\right]^2$ differs from 
that of $g_n^2$.
This difference is not trivial as $\left[c_d(n)\right]^2$ has a dramatic
$\la$ dependence (see Fig. (1) in Ref. \cite{Ab 99b}).
The inclusion of the factor $\f{1}{N}$ in Eq. (\ref{golden2}) makes clear that
\emph{model A} and \emph{model B} are only comparable when $V^2_{Jd}$
is of the order $Nv^2_J$.

\emph{Model A} is precisely equivalent to that of Refs.\cite{We 98,Gu 99}
when $\la=1$. The real part of
Eq. (\ref{ourmatrix}) used in \emph{model B} is equivalent to 
what is used in
Refs. \cite{Ab 99,Ab 99b}, however, 
we calculate the average intensity integrated
over the energy: $F_J^{\rm{av.}}$,
whereas Refs. \cite{Ab 99,Ab 99b} calculate a tunnelling
probability which is more closely related to $W_{JJ}(E)$.

Note that $F_J^{\rm{av.}}$ can be written as
\be
F_J^{\rm{av.}}=\f{\Gamma^{\gamma}_{J}}{2\pi}
\int_{-\infty}^{\infty}dE\f{1}
{(E-E_J-\Delta^{\da}_{J}(E))^2+\f{\Ga^{\ga2}_J}{4}
(1+\f{\Gamma^{\da}_{J}(E)}{\Ga^{\ga}_J})^2},
\label{2FJ}
\ee
where
\be\label{Del}
\Delta^{\da}_{J}(E)=\mbox{Re}W_{JJ}(E),
\ee
and
\be\label{Gam}
\Gamma^{\da}_{J}(E)=-2\mbox{Im}W_{JJ}(E).
\ee
Ignoring the shift $\Delta^{\da}_{J}(E)$ altogether and assuming that
the width $\Gamma^{\da}_{J}(E)$ has the energy independent value $\Ga^{\da}_J$
one obtains the principal result of Ref. \cite{We 98} that
\be\label{3FJ}
F_J^{\rm{av.}}\approx\f{\Gamma^{\gamma}_{J}}{\Gamma^{\gamma}_{J}+\Ga^{\da}_J}.
\ee

\begin{figure}[ht]
\centerline{\psfig{figure=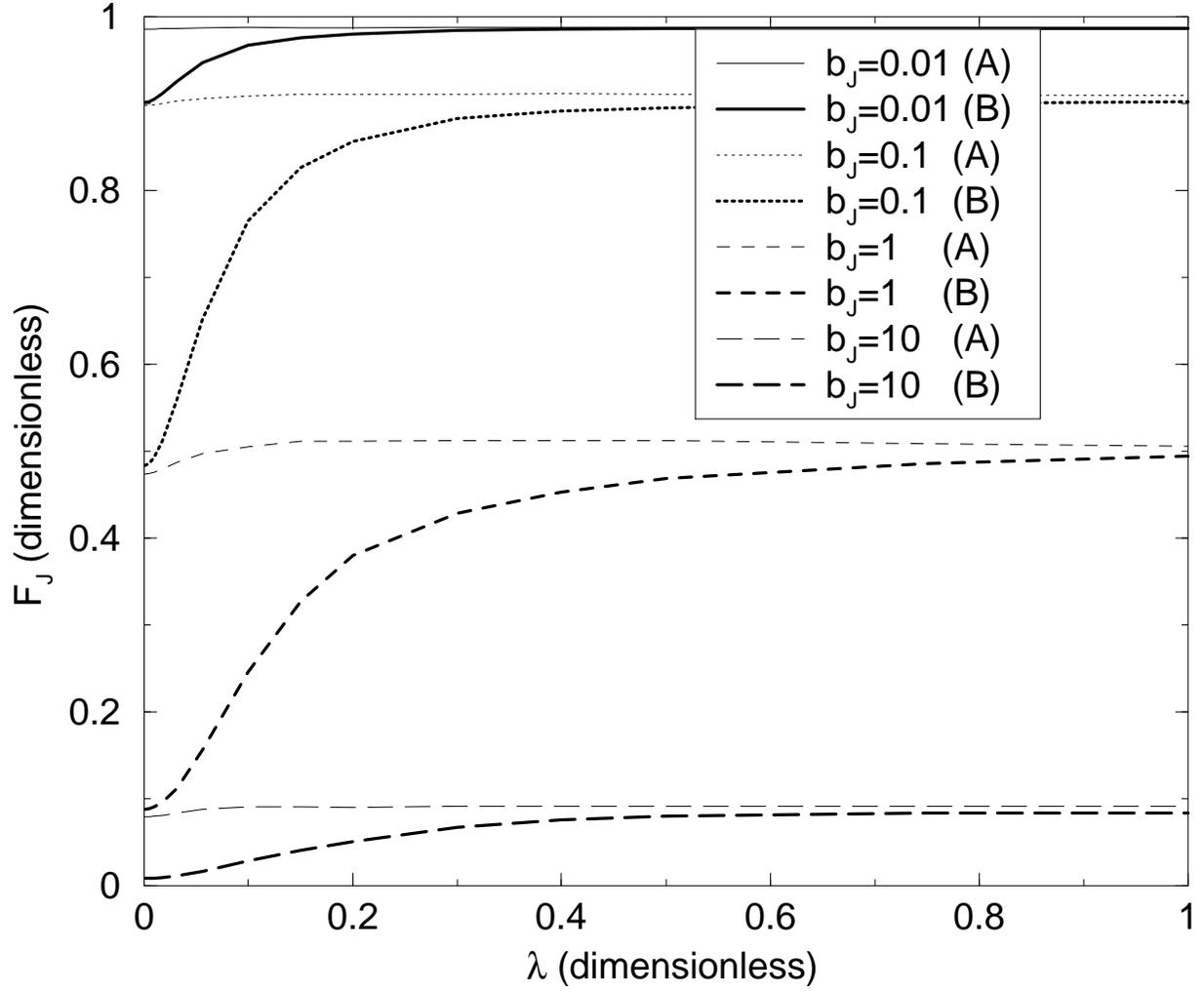}}
\caption{Calculated attenuation factor $F_J^{\rm{av.}}$ 
as a function of $\la$ for some values of 
$b_J=\f{\Gamma^{\da}_{J}}{\Gamma^{\gamma}_{J}}$. The thin
lines where calculated using \emph{model A} and the thick lines using
\emph{model B}.} 
\label{FJfig1lin}
\end{figure}

\begin{figure}[ht]
\centerline{\psfig{figure=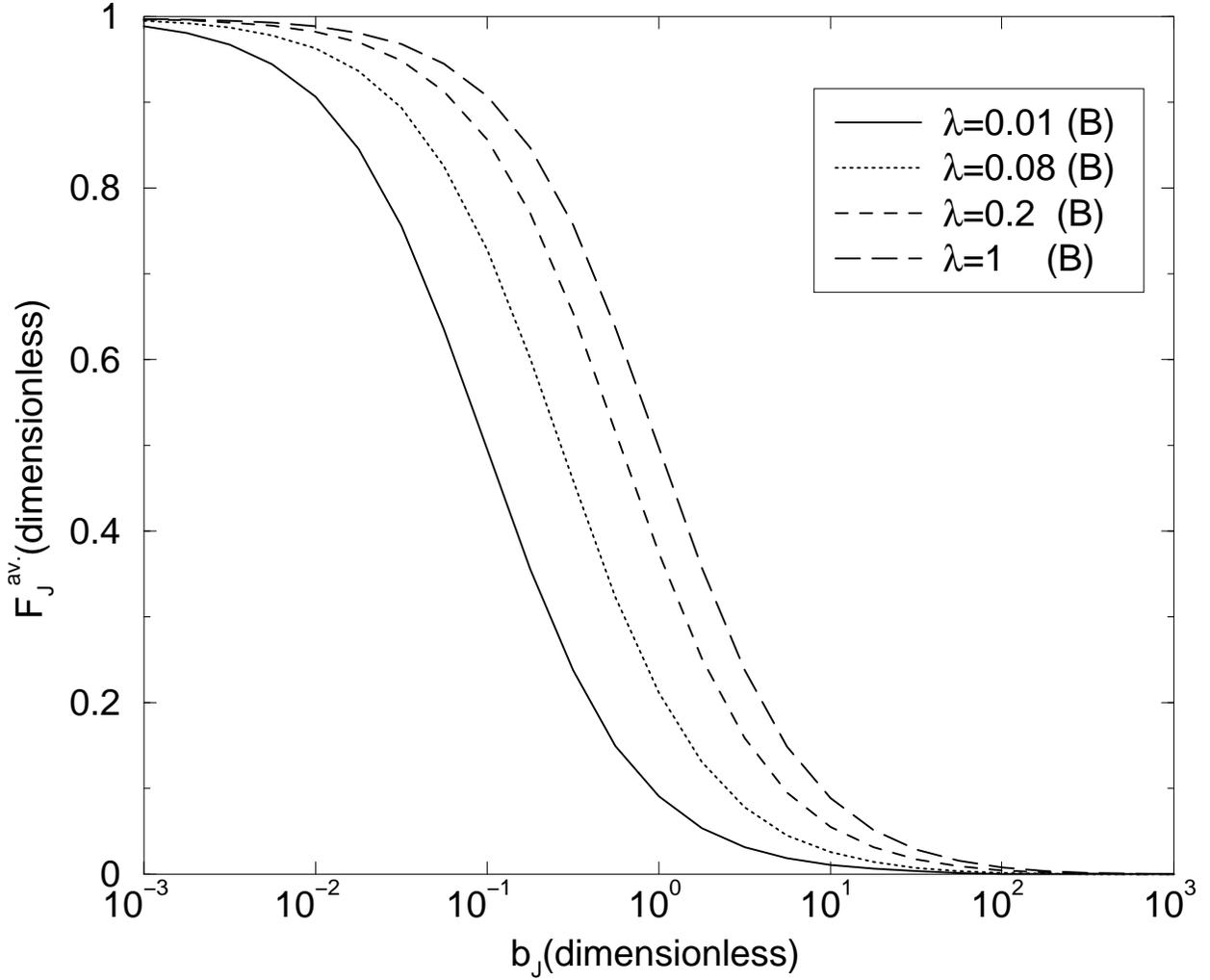}}
\caption{Attenuation factor $F_J^{\rm{av.}}$ 
as a function of $b_J=\f{\Gamma^{\da}_{J}}{\Gamma^{\gamma}_{J}}$ for some
values of $\la$, calculated using \emph{model B}.}
\label{FJfig2}
\end{figure}

We now present numerical calculations of $F_J^{\rm{av.}}$ with $N=50$,
$\Ga^{\ga}_J=0.01 D_J$. An ensemble average
was performed over 100 realisations in Fig. (\ref{FJfig1lin}) 
and over 1000 realisations in Fig (\ref{FJfig2}).
The effect of increasing $\Ga^{\ga}_N$, identical
to that obtained by increasing the energy averaging interval $I$, is to
broaden $\Gamma^{\da}_{J}(E)$ (Eq. (\ref{Gam})) and
thus push $F_J^{\rm{av.}}$ closer to the
approximation given by Eq. (\ref{3FJ}). This is in line with what it is 
reported in \cite{St 99} who obtain Eq. (\ref{3FJ}) in the limit 
$\f{\Gamma^{\ga}_{N}}{\Gamma^{\ga}_{S}}\rightarrow \infty$ for their two
level model. 

In our calculations we put $\Ga^{\ga}_N+I=3D_J$. With this choice one may 
describe what Ref. \cite{Gu 99} calls the overlapping resonance region.
Ref. \cite{Gu 99} gives the impression that the relative intensity is 
independent of $\Gamma^{\ga}_{J}$. Whilst we agree 
with \cite{Gu 99} that the ratios
$b_J=\f{\Gamma^{\da}_{J}}{\Gamma^{\gamma}_{J}}$ and $\f{\Ga^{\ga}_N}{D}$ 
are of 
principal importance in understanding the decay out, it can be seen
from Eq. (\ref{2FJ}) that $F_J^{\rm{av.}}$ is only independent of 
$\Gamma^{\ga}_{J}$ if $\Delta^{\da}_{J}(E)$ and $\Ga^{\da}_{J}(E)$
are constant.

Fig. \ref{FJfig1lin}
shows the dependence of $F_J^{\rm{av.}}$ on $\la$, the 
strength of the
mixing amongst the ND states, for several values of 
$b_J$, for both \emph{model A} and \emph{model B}.
For \emph{model A} the variation of $F_J^{\rm{av.}}$
with $\la$ is rather slight compared to \emph{model B}.
This is because the $\la$
dependence of \emph{model A} is contained in the eigenvalues $E_n$
which are unfolded to have unit mean spacing. \emph{Model B}
has a further and more significant $\la$ dependence contained in the
eigenvectors $c_d(n)$.
For \emph{model B}, $F_J^{\rm{av.}}$
decreases with decreasing $\la$ to a value which is limited by the value
of $\Ga^{\ga}_N+I$. Note that $F_J^{\rm{av.}}$
can change at most by a factor of about 5 by varying $\la$. 

Fig. \ref{FJfig2} shows $F_J^{\rm{av.}}$ as function of $b_J$ for some
values of $\la$, calculated using \emph{model B}. The calculations for 
\emph{model A} are not shown as they can barely be distinguished from the
calculation for $\la=1$ using \emph{model B}.
The effect of changing $\la$ is to move the value
of $b_J$ (and hence $J$) at which the decay out occurs. Thus from 
Fig. \ref{FJfig2} we conclude that the decay out is slightly
hindered by increasing $\la$. 

Regarding Refs. \cite{Ab 99,Ab 99b} which report an increase in the
tunnelling probability of several orders of magnitude with increasing 
$\la$ we do not consider ourselves at odds with this work since, as
already mentioned above, we do not calculate the same quantity. A further
difference between \emph{model B} of this paper and 
Refs. \cite{Ab 99,Ab 99b} is that their author
places $|d\ket$ at the position
$\f{N}{4}$ thus making the distribution $c_d(n)$ asymmetric. This 
would correspond
in our calculation to making the difference $E_J-E_d$ non-zero
(we see no reason not to set $E_J=E_d$).

An investigation of the
roles of $\Ga^{\ga}_N$ and $I$ we postpone to a subsequent paper. The results
of Ref \cite{Gu 99} indicate that $F_J^{\rm{fluc.}}$, Eq. (\ref{Ffl}),
is important when $\Ga^{\ga}_N$ is a small fraction of $D_J$ 
(non-overlapping resonance region).

It was already found from phenomenological analysis  
some years ago \cite{Vi 90b,Vi 90a,Kh 93} that
$F_J^{\rm{av.}}$ falls exponentially with decreasing spin. We conclude here 
that the chaotic nature of the ND states, as classified by $\la$, 
cannot account for such behaviour.
The exponential drop in the intra-band intensity 
must be due to the spin dependence of the tunnelling matrix element 
contained in $b_J$. The calculation of $b_J$ is not trivial and we
refer the reader to \cite{Sh 00,Yo 00},
which continue the work of Refs. \cite{Vi 90b,Vi 90a,Sh 92,Sh 93}, for
some recent calculations.

A.J.S. thanks J.A. Tostevin for his comments on an early version
of this paper. This work was supported by FAPESP.


\end{document}